\documentclass[
 reprint,
 amsmath,
 amssymb,
 aps,
 pra, %
 superscriptaddress,
]{revtex4-2}

\usepackage[utf8]{inputenc}
\usepackage{graphicx}

\usepackage{gensymb}
\usepackage{siunitx}
\usepackage{threeparttable, tablefootnote}
\usepackage{booktabs}
\usepackage{csquotes}
\usepackage[export]{adjustbox}
\usepackage{placeins}
\usepackage{setspace}
\usepackage{enumitem}
\usepackage{verbatim}
\usepackage{mathtools}
\usepackage[thinc]{esdiff}
\usepackage{amsmath}
\usepackage[caption=false]{subfig}
\usepackage[usenames,dvipsnames]{color}
\usepackage{braket}

\usepackage{dcolumn}
\usepackage{bm}
\usepackage[utf8]{inputenc}
\usepackage[T1]{fontenc}
\usepackage{mathptmx}
\usepackage{etoolbox}
\usepackage{hyperref}
\usepackage{cleveref}

\newcommand{\ion}[2]{\ensuremath{^{#2}\mathrm{#1}^+}}
\newcommand{\hfket}[3]{\ensuremath{\ket{\mathrm{#1},\mathrm{F}={#2}, \mathrm{m}_\mathrm{F}={#3}}}}

\newcommand{\MITAffiliation}[0]{Center for Ultracold Atoms, Research Laboratory of Electronics, Massachusetts Institute of Technology, Cambridge, Massachusetts 02139, USA}
\newcommand{\LLAffiliation}[0]{Lincoln Laboratory, Massachusetts Institute of Technology, Lexington, Massachusetts 02421, USA}

\begin{document}

\title{Spontaneous Raman scattering from metastable states of Ba$^+$}

\author{Timothy J. Burke}
\thanks{These two authors contributed equally}
\email{tburke2@mit.edu}
\affiliation{\MITAffiliation}

\author{Xiaoyang Shi}
\thanks{These two authors contributed equally}
\affiliation{\MITAffiliation}

\author{Jasmine Sinanan-Singh}
\affiliation{\MITAffiliation}

\author{Isaac L. Chuang}
\affiliation{\MITAffiliation}

\author{John Chiaverini}
\affiliation{\MITAffiliation}
\affiliation{\LLAffiliation}

\begin{abstract}
Quantum logic gates performed via two-photon stimulated-Raman transitions in ions and atoms are fundamentally limited by spontaneous scattering errors. Recent theoretical treatment of these scattering processes has predicted no lower bound on the error rate of such gates when implemented with far-detuned lasers, while also providing an extension to metastable qubits. To validate this theoretical model, we provide experimental measurements of Raman scattering rates due to near-, and far-detuned lasers for initial states in the metastable D$_{5/2}$ level of \ion{Ba}{137}. The measured spontaneous Raman scattering rate is consistent with the theoretical prediction and suggests that metastable-level two-qubit gates with an error rate $\approx10^{-4}$ are possible with laser excitation detuned by tens of terahertz or more.
\end{abstract}

\maketitle

\section{Introduction}

    Stimulated Raman transitions are essential operations for many Rydberg-atom~\cite{Harry2022} and trapped-ion~\cite{Ballance2016} systems, as they allow for high-fidelity entangling quantum gates~\cite{gaebler_high-fidelity_2016}, sub-Doppler cooling~\cite{Kerman2000}, and high-resolution spectroscopy~\cite{Berto2014}. Errors in quantum-logic gates in trapped-ion systems carried out via such transitions have been reduced to the level where fundamental limits due to spontaneous Raman scattering (SRS) dominate~\cite{Ballance2016, gaebler_high-fidelity_2016}. SRS is a source of decoherence in which a spontaneously scattered photon carries information of the internal atomic state, effectively measuring it. To characterize and improve Raman-based two-qubit gate fidelity, it is crucial to have an accurate description of SRS.


    The recently proposed \textit{omg}/dual-type encoding methodology~\cite{allcock_omg_2021, yang_realizing_2022} provides new capabilities when utilizing metastable states in combination with ground-state encodings in a single-species ion or atom system.  The large energy separations between manifolds housing metastable-state and ground-state qubits naturally lead to a separation of functions, allowing for low-error quantum information processing.  Experimental implementations utilizing metastable qubits in this paradigm have included demonstrations of error correction schemes~\cite{debry2025errorcorrectionlogicalqubit,quinn2024highfidelityentanglementmetastabletrappedion,debry_experimental_2023}, reduced qubit cross-talk \cite{vizvary_eliminating_2024}, high-fidelity SPAM \cite{sotirova2024highfidelityheraldedquantumstate}, and two qubit gates \cite{bazavan_synthesizing_2023, wang_experimental_2025}. Stimulated Raman transitions can supply an optically addressable method for operations within the metastable manifold, and so it is important to understand the SRS process in metastable qubits.

    \begin{figure}[h]
        \centering
        \includegraphics[width=0.99\columnwidth]{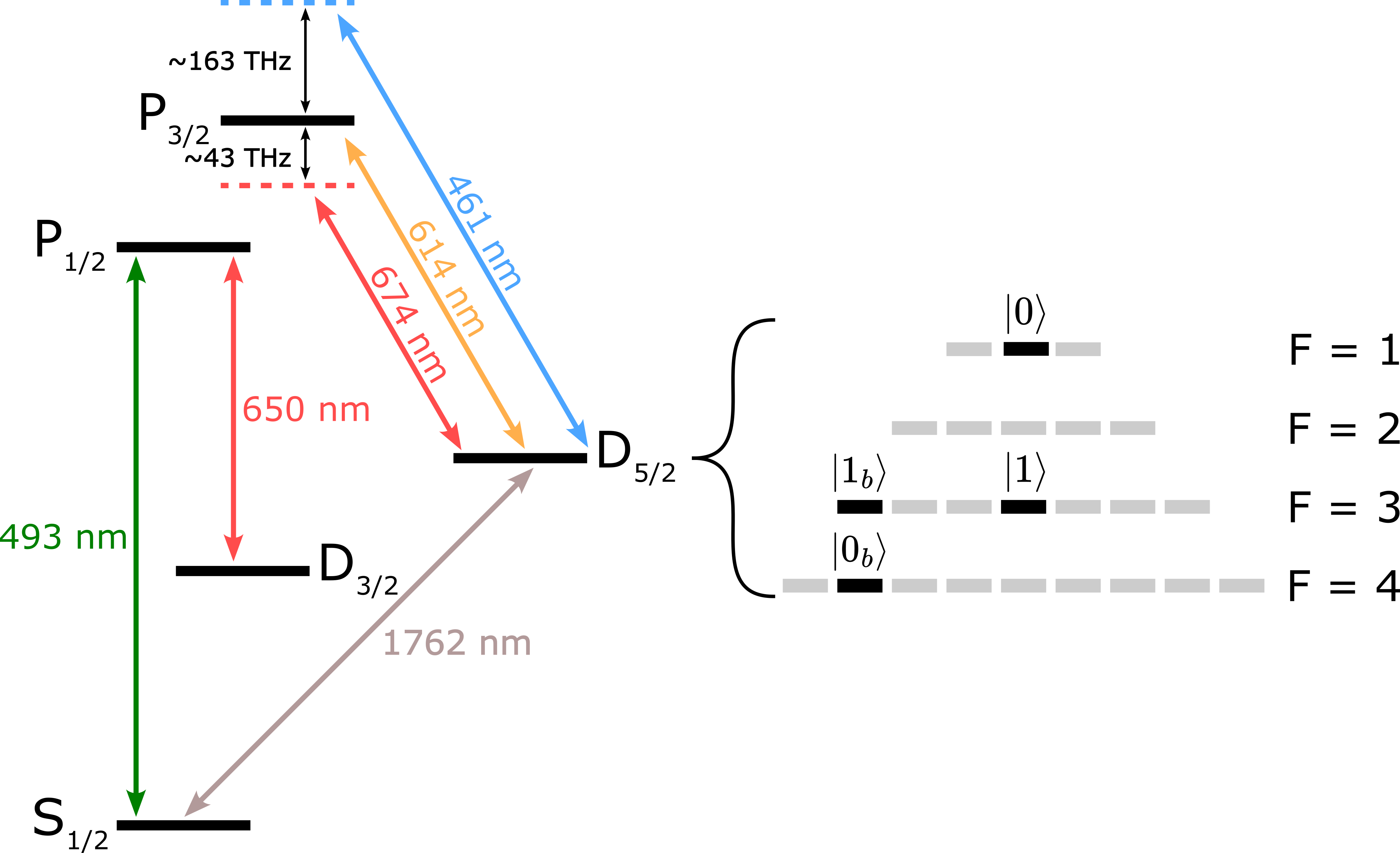}
        \caption{The level structure and lasers used for the experiment. The 493~nm and 650~nm wavelength fields are the Doppler and repump lasers, respectively,and the 1762~nm field drives the quadrupole transition to the metastable D$_{5/2}$ level. The 674~nm, 617~nm, and 461~nm lasers are used for the scattering measurement. The D$_{5/2}$ manifold is shown with relevant states, where $\ket{0}=\hfket{5D_{5/2}}{1}{0}$ and $\ket{1}=\hfket{5D_{5/2}}{3}{0}$ are the qubit states used, with scattering being measured out of $\ket{0}$. While the ``best'' qubit, $\ket{0_b}=\hfket{5D_{5/2}}{4}{-3}$ and $\ket{1_b}=\hfket{5D_{5/2}}{3}{-3}$, is the metastable qubit with the minimum SRS rate.}
        \label{fig:levels}
    \end{figure}

    SRS error in ground state qubits was described by Ozeri et al.~\cite{ozeri_2005} leading to the prediction of decreasing scattering rates as laser detuning increases, but only to a specific lower bound, putting a limit on the achievable operation fidelity regardless of detuning. On the other hand, a more detailed treatment~\cite{moore_2023} suggests that the SRS rate can be arbitrarily lowered with further red-detuning of the applied laser fields. Experimental measurements \cite{boguslawski_2023} using high intensity, red-detuned light on a ground state qubit in \ion{Ba}{133} is in good agreement with this updated theory, and the measured error rates are lower than the bound predicted in ~\cite{ozeri_2005}. However, experimental validation of the SRS theory has not been performed for metastable qubits. The difference between the two scattering models is small for small detunings, even more so for metastable qubits than for ground-state qubits. Hence, differentiating the models requires measurements at either very high laser power in order to gain the necessary statistics in reasonable time frames, or at large detunings where the models diverge.

   In this work, we provide experimental results supporting the theoretical model of \cite{moore_2023} with near-resonant, red-detuned, and blue-detuned SRS measurements in metastable \ion{Ba}{137} qubits (see level structure in Fig.~\ref{fig:levels}). We show that two-qubit gate errors due to SRS can potentially be reduced to the $10^{-4}$ level with far-detuned light and reasonable gate times.

   Below, we briefly describe the scattering model proposed by \cite{moore_2023} and the theoretical tools used in the course of the measurement. We then describe our experimental technique and how we extract the SRS rate from our measurements. We conclude with a comparison between the two scattering models and a discussion of achievable two-qubit gate errors in the metastable D$_{5/2}$ level of Ba$^{+}$.
   Pedagogical mathematical derivations are left for the appendix.

\section{Theory}

\begin{figure}[tbp]
        \centering
        \includegraphics[width=0.5\textwidth]{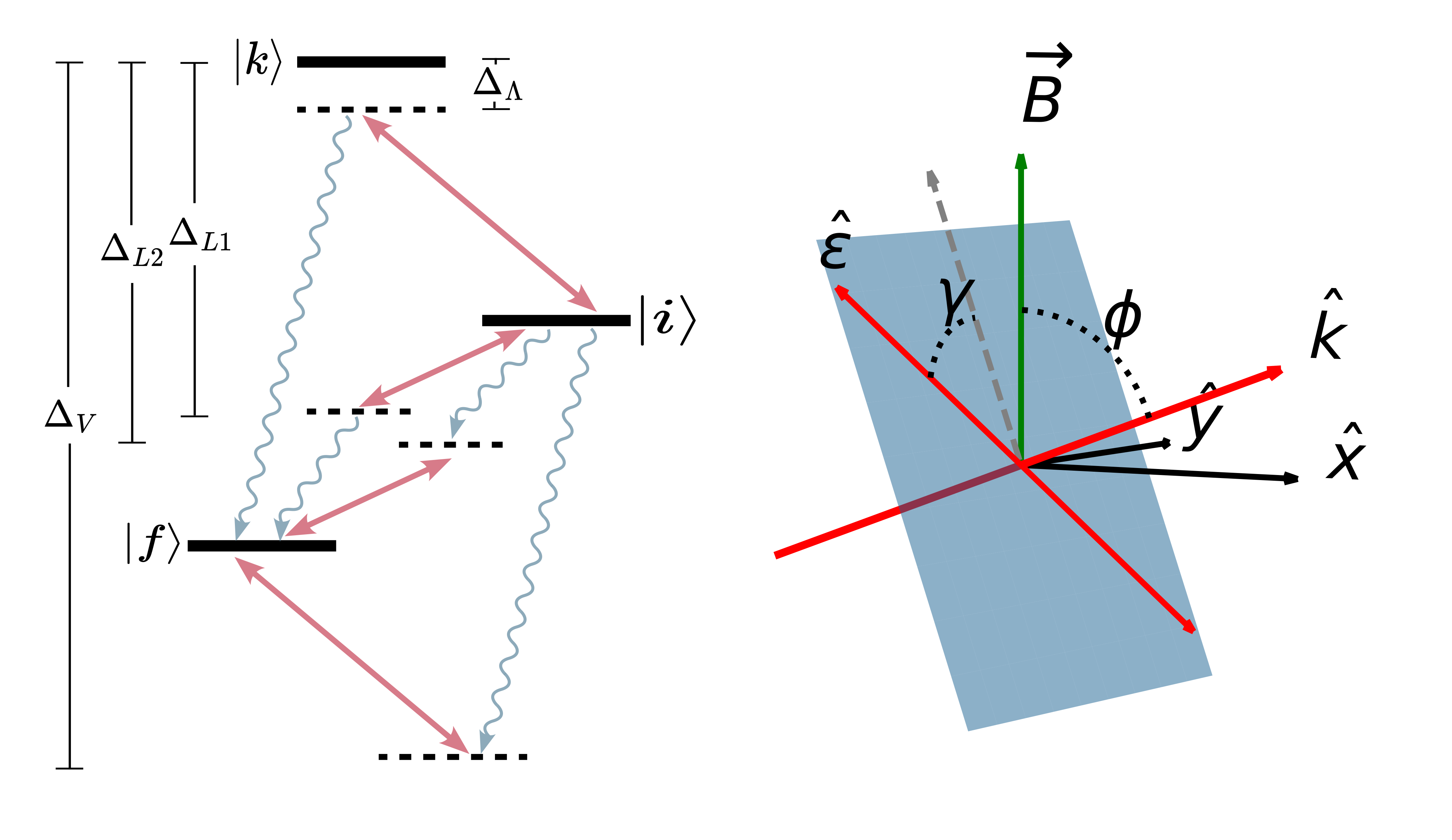}
        \caption{Relevant scattering processes and experimental beam geometry. (left) The energy levels involved in scattering are depicted, where $\ket{i}$ and $\ket{f}$ are the initial and final states, and $\ket{k}$ is the intermediate state that is scattered from. Each scattering process ($\Lambda$, $V$, and the two ladder processes) has a corresponding detuning ($\Delta_\Lambda$, $\Delta_V$, $\Delta_{L1}$, and $\Delta_{L2}$), though $\Delta_\Lambda$ is the quantity being referred to when ``detuning'' is used without further specification. The laser field is shown in red while the scattered photon is blue. (right) Perspective diagram of the scattering beam. The laser-beam direction $\hat{k}$ is in the $\hat{x}$-$\hat{z}$ plane with angle $\phi$ to the quantization axis, $\overrightarrow{B}$, which is aligned with $\hat{z}$. The polarization vector $\hat{\epsilon}$ is necessarily in the plane normal to $\hat{k}$ and is parameterized by angle $\gamma$ to the quantization axis projected into the plane. }
        \label{fig:scat_procs}
    \end{figure}

    SRS can be thought of as off-resonant scattering from some rapidly decaying intermediate state(s) $\ket{k}$. In a recently suggested model~\cite{moore_2023}, four scattering processes are considered, as shown in Fig.~\ref{fig:scat_procs}. Starting from the initial state $\ket{i}$, a laser detuned from the intermediate state $\ket{k}$ couples the two states. During the Raman process, an SRS event could occur, in which a photon is first absorbed and then emitted, or these events could take place in reverse order, leaving the ion in the final state $\ket{f}$. We label the two processes $\Lambda$ and $V$ scattering, respectively. Additionally, if the laser frequency is less than the transition frequency between $\ket{i}$ and $\ket{f}$, two additional scattering processes could occur in which a photon is emitted into the laser field through stimulated emission and another photon is emitted spontaneously; the order in which the photons are emitted differentiates these two ``ladder'' processes. Each of these processes is facilitated by coupling to intermediate states through electric dipole interactions.
    
    The predicted SRS rate from $\ket{i}$ to $\ket{f}$ is given by \cite{moore_2023}:
    \begin{widetext}
    \begin{align}
        \Gamma_{i,f} &=\Theta(\omega_{sc, \Lambda V}) \frac{E_\ell^2 \omega_{sc, \Lambda V}^3}{12 \pi \varepsilon_0 \hbar^3 c^3} \sum_q \left| \sum_{k} \left(\frac{\langle f | \bm{r} \cdot \hat{\epsilon}_q | k \rangle \langle k | \bm{r} \cdot \hat{\epsilon}_\ell | i \rangle}{\Delta_\Lambda} +\frac{\langle f | \bm{r} \cdot \hat{\epsilon}_\ell | k \rangle \langle k | \bm{r} \cdot \hat{\epsilon}_q | i \rangle}{\Delta_V} \right) \right|^2 
        \label{eq:scatter} \\ 
        &+\Theta(\omega_{sc, L}) \frac{E_\ell^2 \omega_{sc, L}^3}{12 \pi \varepsilon_0 \hbar^3 c^3} \sum_q \left| \sum_{k} \left(\frac{\langle f | \bm{r} \cdot \hat{\epsilon}_q | k \rangle \langle k | \bm{r} \cdot \hat{\epsilon}_\ell^* | i \rangle}{\Delta_{L1}} +\frac{\langle f | \bm{r} \cdot \hat{\epsilon}_\ell^* | k \rangle \langle k | \bm{r} \cdot \hat{\epsilon}_q | i \rangle}{\Delta_{L2}} \right) \right|^2 \nonumber
    \end{align}
    \end{widetext}
    
        \noindent where $E_\ell$ is the electric field amplitude of the lasers, $\bm{r}$ is the dipole moment, and the detuning for each process is defined as $\Delta_\Lambda=(E_k-E_i)/\hbar-\omega_\ell$, $\Delta_V=(E_k-E_f)/\hbar+\omega_\ell$, $\Delta_{L1}=(E_k-E_i)/\hbar+\omega_\ell$, and $\Delta_{L2}=(E_k-E_f)/\hbar-\omega_\ell$.  Here $E_i$, $E_k$, and $E_f$ are the energies of the initial, intermediate, and final states, respectively. The quantity $\hbar \omega_{sc, \Lambda V}= E_k-\hbar\Delta_\Lambda-E_f = E_i+\hbar\Delta_V-E_k$ is the energy of the scattered photon for $\Lambda$ or $V$ scattering, $\omega_\ell$ is the frequency of the laser, and $\hbar\omega_{sc, L}=E_k-\hbar\Delta_{L1}-E_f=E_i+\hbar\Delta_{L2}-E_k$ is the energy of the scattered photon for ladder scattering. We also define $\hat{\epsilon}_q$ as the scattered photon polarization unit vector, and $\hat{\epsilon}_\ell$ is the laser polarization unit vector. The Heaviside function $\Theta(\cdot)$ enforces energy conservation, the sum over $q$ accounts for all polarizations since the scattered photon is spontaneous, and the sum over $k$ includes all common states between $\ket{i}$ and $\ket{f}$ that are connected by dipole-allowed transitions. The four terms correspond to each scattering process shown in Fig.~\ref{fig:scat_procs}: $\Lambda$, V, and two ladder scattering channels. We use reduced matrix elements and lifetimes from \cite{UDportal}, and energies from \cite{NIST_ASD}. Previous work by Ozeri et al.~\cite{ozeri_2005} describes a simpler model for the scattering rates from ground states which gives accurate results for small detuning. We extend the model to metastable states in the Appendix so that a comparison can be made with the model provided by Moore et al.~\cite{moore_2023}; see \cref{eq:ozscatter}.

\section{Experiment}
    \begin{figure*}
        \centering
        \includegraphics[width=\textwidth]{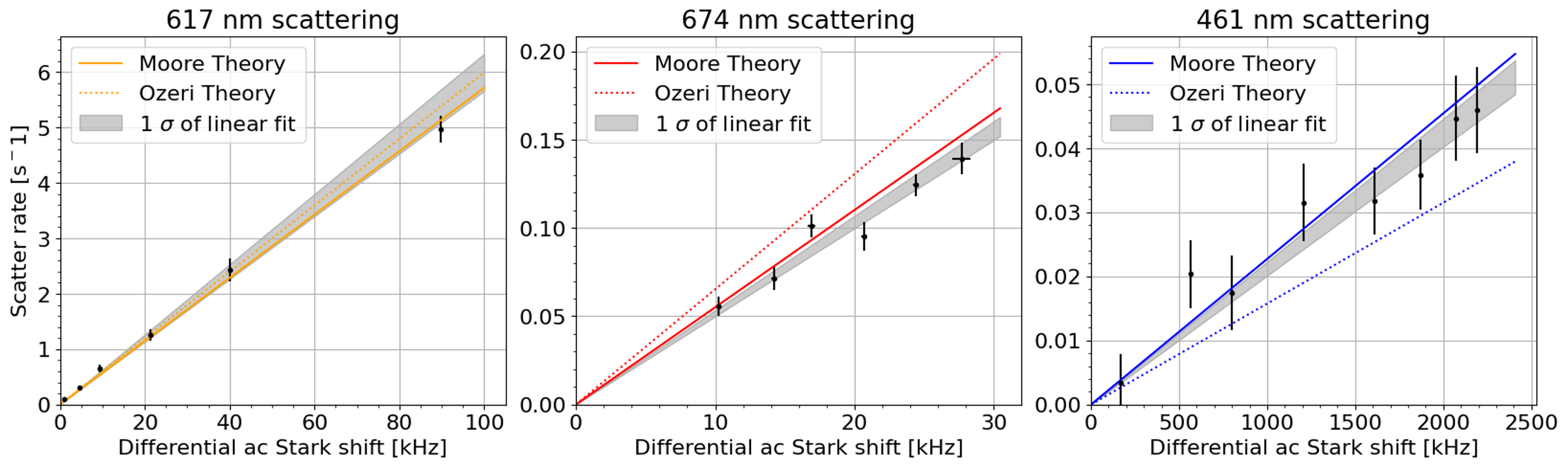}
        \caption{Experimental measurement of SRS rate into $6S_{1/2}$ and $5D_{3/2}$ vs. differential AC Stark shift for 617, 674, and 461 nm, respectively. The solid black line represents a linear fit to the data points and the shaded region represents the 1 $\sigma$ error from fitting. In order to compare, we plot the theoretical models as $\sum_f\Gamma_{0,f}/\Delta_{d,s}$ (see \cref{eq:scatter}, \cref{eq:ozscatter}, and \cref{eq:diffacstark}) where $\ket{0}=\hfket{5D_{5/2}}{1}{0}$, and the sum over $f$ includes all states in $6S_{1/2}$ and $5D_{3/2}$.}
        \label{fig:scattering_rate_combined}
    \end{figure*}

     We perform the scattering-rate measurement at three wavelengths, as shown in Fig.~\ref{fig:levels}.  We use 461 nm and  674 nm, corresponding to large blue and red detuning, relative to the resonant transition at 614 nm. We also measure the scattering rate at 617 nm to test the validity of the models at small detuning. 
     
     A single \ion{Ba}{137} is trapped in a cryogenic surface-electrode trap (a more detailed description of the experimental system can be found in \cite{shi_long-lived_2025}) to measure the spontaneous Raman scattering rate. Multiple rounds of state preparation, shelving, and detection are performed to initialize the ion in the desired state \cite{an_high_2022}. The initial state is chosen to be $\ket{0}=\hfket{5D_{5/2}}{1}{0}$ as this maximizes the scattering rate. The scattering light is then switched on, and we check, every 100 ms (10 ms for 617 nm due to the higher scattering rate), to see if the ion has decayed to the S$_{1/2}$ or D$_{3/2}$ state by measuring the fluorescence from the ion during application of the Doppler and repumping laser beams at 493~nm and 650~nm, respectively. The decay rate is determined by fitting the probability of the ion not fluorescing at different delay times to an exponentially decaying function. The scattering rate is extracted from the difference between the measured lifetime with scattering light versus auxiliary measurements made of the natural lifetime (without scattering light) of the D$_{5/2}$ state. For the measurement at 461 nm, since the light would also significantly couple the 6S$_{1/2}$ state to the 6P$_{3/2}$ state, which could decay to the 5D$_{5/2}$ state, we apply a low-intensity 493 nm laser beam to the ion to pump any decayed population to the 5D$_{3/2}$ state. Due to the larger detuning and relatively lower intensity, the 493 nm light has negligible impact on the measured scattering rate. The measurement is repeated at different laser intensities; the intensity is determined by measuring the differential AC stark shift between the $\hfket{6S_{1/2}}{1}{-1}$ and $\hfket{5D_{5/2}}{1}{1}$ states.

    To measure the laser-beam polarization, we compare measurements of the Rabi frequencies of different transitions. For all three wavelengths, the polarization of the scattering light is set to be linear and parallel with the magnetic field $\vec{B}$, and the $k$-vectors of the laser beams are nominally perpendicular to $\vec{B}$. To make these auxiliary measurements with the 674-nm light, we trap a single \ion{Sr}{88} ion in the same trap, with no \ion{Ba}{137} ion present, and drive transitions between the 5S$_{1/2}$ and 4D$_{5/2}$ levels with different net changes in the magnetic quantum number $\Delta m_J$. We measure $\Omega_{\Delta m_J=0} << \Omega_{\Delta m_J=1,2}$ and therefore can assume that the laser $k$-vector is perpendicular to the quantization axis (see fig. 3.9 in \cite{Roos2000} or A.6 from \cite{James_1998}). We then perform a least-squares fit of \cref{eq:e2rabi} to $\Omega_{\Delta m_J=1,2}$ to extract the polarization (angle $\gamma$; see Fig.~\ref{fig:scat_procs}). 
    
    For the measurements at 617 nm, two-photon stimulated-Raman transitions are driven between the metastable states in \ion{Ba}{137} to extract the polarization by comparing the Rabi frequencies in a similar way. We can assume, from optomechanical constraints, that the $k$-vectors of the two 617-nm Raman beams are perpendicular to each other and have the same angle with respect to the quantization axis (the beams are directed to the ion through perpendicularly positioned windows with angles just large enough to avoid scattering light from two perpendicular edges of the square trap chip). We then fit to \cref{eq:ramanrabi}, parameterized by the two beam polarizations and wavevector angles. See Table~\ref{tab:pols} for the results of the fits determining the 674-nm and 617-nm beam polarizations.
    
    For the auxiliary measurement with the highly detuned 461-nm laser beam, we could potentially make a similar measurement with Raman transitions in the ground state of \ion{Ba}{137}, but $\Delta m=2$ transitions are not possible with linearly polarized beams in the $S_{1/2}$ manifold (see~\cref{eq:ramanrabi}). Therefore, we validate that the polarizations are similar to those of the 617-nm beams. This is a good assumption, as both the 617-nm and 461-nm beams enter the vacuum chamber through the same beam path and thus have the same wavevectors. To confirm this assumption, we use a polarizing beam splitter and a power meter to measure the polarization of the transmitted beam. To eliminate a possible confounding effect caused by the wavelength dependence of birefringence in the chamber windows, we measure the degree to which the polarization is changed upon transmission through the windows at both 617~nm and 461~nm. We find a negligible difference. Thus, we take the polarization impurity for the 461-nm measurement to be equivalent to that of the 617-nm measurement.

    \begin{table}
    \begin{tabular}{|c|c|}
    \hline
        $\lambda$ [nm] & Polarization angle [rad] \\
        \hline
        674 & 0.045 $\pm$ 0.001 \\
        617 & 0.105 $\pm$ 0.006 \\
        \hline
    \end{tabular}
    \caption{Experimentally determined polarizations of the beams used for scattering measurement, with standard errors derived from least squares fits to the Rabi frequencies. The 461 nm scattering-measurement analysis uses the polarization determined from the 617 nm, see text.}
    \label{tab:pols}
    \end{table}

    \begin{figure*}
    \centering
    \includegraphics[width=\textwidth]{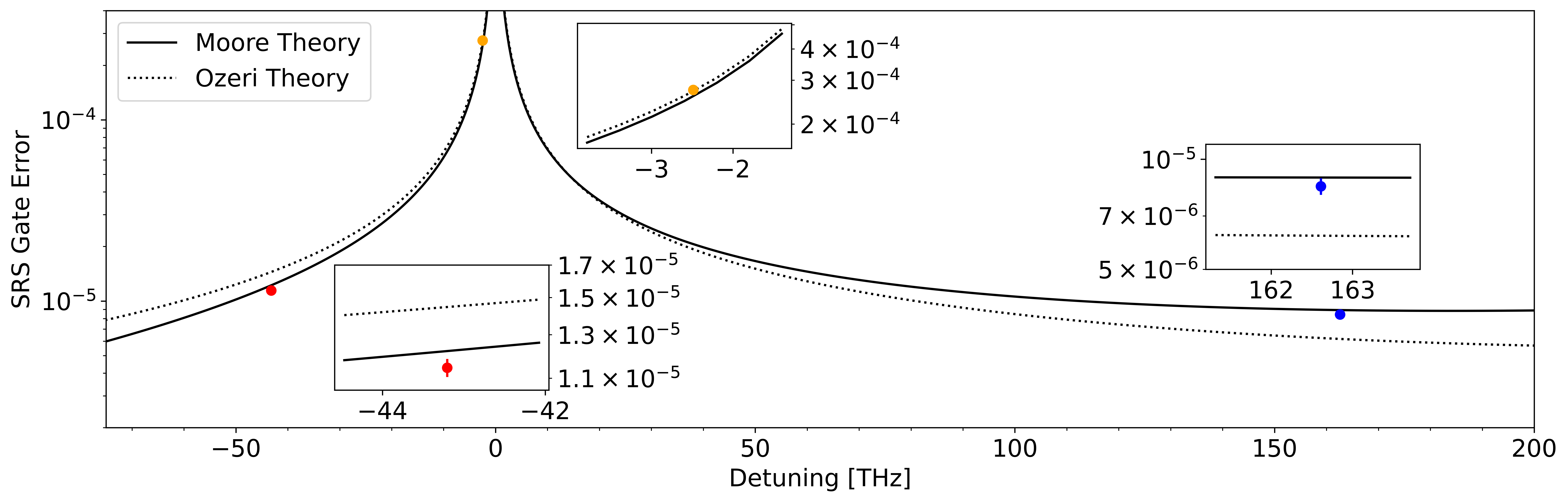}
    \caption{Representative gate error due to SRS during a single qubit, $\sigma_x$ Raman gate as a function of laser detuning from the $5D_{5/2}$-$6P_{3/2}$ transition. More specifically, plotted is $\Gamma_{\ket0}\tau_\pi$, the scatter probability from $\ket{0}$ into any S$_{1/2}$ or D$_{3/2}$ states during a $\pi$-pulse from $\ket{0}=\hfket{5D_{5/2}}{1}{0}$ to $\ket{1}=\hfket{5D_{5/2}}{3}{0}$; this allows for comparison to the experimental data points corresponding to the three measurements shown in Fig.~\ref{fig:scattering_rate_combined}. The actual SRS gate error for this operation would be $\frac12(\Gamma_{\ket0} + \Gamma_{\ket1})\tau_\pi$, since both states are populated, on average, for half of the duration of the gate. Also, scattering back into D$_{5/2}$ (excluding $\ket{0}$) would need to be be included.}
    \label{fig:1q_err}
    \end{figure*}

\section{Discussion and Conclusion}
    The results of our measurements are shown in Fig.~\ref{fig:scattering_rate_combined}. For each detuning, we plot SRS rates versus measured differential AC Stark shift. We plot the predicted SRS rate (\cref{eq:scatter}) for both theories and calculate the expected AC Stark shift (\cref{eq:diffacstark}), as this allows for a direct comparison. We see better agreement with the Moore theory, limited by measurement statistics due to the low scatter rates at large detunings. (Our measurements cannot distinguish between the theories at small detuning, where the measurements agree with both, within uncertainty.)


    Shown in Fig.~\ref{fig:1q_err} is a representative single-qubit gate error for a $\sigma_x$ two-photon stimulated-Raman gate operation. The experimental points are derived from the measured scattering rate by multiplying the slope (scatter rate per Stark shift) of the linear fits shown in Fig.~\ref{fig:scattering_rate_combined} by the calculated differential AC Stark shift (\cref{eq:diffacstark}) and gate time (\cref{eq:ramanrabi}). Notably, if we assume that the $\sigma_x$ gate implemented with 617-nm beams has a typical gate time of 5 $\mu$s, using 674-nm or 461-nm beams with the same total intensity will result in gate times of 80~$\mu$s or 484 $\mu$s, respectively.

    We note that the contribution from the ladder terms in \cref{eq:scatter} is not tested since the energy conservation condition is not met (ladder scattering can only occur if $(E_i-E_f)>\hbar\omega_\ell$). Our choice of qubit state $\ket{0}$ was made in order to maximize scattering to achieve more statistically significant results; this also means the gate-error values shown in Fig.~\ref{fig:1q_err} are the worst-case result. For example, a $\sigma_x$ gate with 617-nm beams using the qubit $\ket{0_b}=\hfket{5D_{5/2}}{4}{-3}$ and $\ket{1_b}=\hfket{5D_{5/2}}{3}{-3}$ has a predicted scatter probability of $7.5\times10^{-5}$, compared with the $3\times10^{-4}$ error rate shown here.

    We can extend these results to illustrate typical two-qubit error rates due to SRS during a Raman gate; this is shown in Table~\ref{tab:gateerrs}. We consider the two-qubit, Mølmer-Sørensen (MS) gate~\cite{MS_gate} in the optimal 3-beam configuration, consisting of two co-propagating beams with power $P$ and a third, counter-propagating beam with power $2P$. The two-qubit error rate can be approximated by scaling the single-qubit gate error by a factor of $\frac{4}{\eta}$, where $\eta$ is the Lamb-Dicke parameter, and the factor of 4 accounts for the presence of two ions and the increase in laser power. The two qubit gate times scale similarly as the single qubit gates discussed above. Assuming a 50~$\mu$s duration MS gate implemented with 617 nm beams, gates driven with the same power at 674 and 461 nm will have gate times of 870 $\mu$s and 3.6 ms, respectively. We note that such long gate times may lead to the situation where fidelities are no longer limited by SRS.

    \begin{table}
    \begin{tabular}{|c|c|c|c|}
    \hline
        $\lambda$ [nm] & Scaled experiment & Predicted & Best qubit\\
        \hline
        617 & $2.3(1)\times10^{-2}$ & $2.2\times10^{-2}$ & $4.6\times10^{-3}$\\
        674 & $1.0(1)\times10^{-3}$ & $1.1\times10^{-3}$ & $2.3\times10^{-4}$\\
        461 & $6.2(3)\times10^{-4}$ & $6.6\times10^{-4}$ & $1.4\times10^{-4}$\\
        \hline
    \end{tabular}
    \caption{Two-qubit gate errors at the three measured wavelengths. We use a motional frequency typical of our system of $\omega =2\pi \times 2$~MHz to determine the Lamb-Dicke parameter. The ``Scaled experiment'' are the values shown in Fig.~\ref{fig:1q_err} multiplied by $\frac{4}{\eta}$ and scaled to account for half the population residing in the $\ket{1}$ state and for scatter back into the D$_{5/2}$ manifold (excluding Rayleigh scattering back to $\ket{0}$).  The ``Best qubit'' column shows the predicted error using the best pair of states for minimizing the SRS error ($\ket{0_b}$ and $\ket{1_b}$).}
    \label{tab:gateerrs}
    \end{table}

    In conclusion, we present measurements of SRS scattering due to near-detuned (-2.5 THz), and far-blue- (163 THz) and far-red- (-43 THz) detuned lasers for metastable states of \ion{Ba}{137}, in good agreement with a recently published model by Moore et al.~\cite{moore_2023}. We show that the SRS-induced error can be reduced below $10^{-5}$ per single-qubit gate for both far-red and far-blue detuned beams. Furthermore, results for metastable states in \ion{Ca}{40}~\cite{moore_spontaneous_2025} produced in parallel with the work presented here also support these conclusions.

    This research was supported by the U.S. Army Research Office through grant W911NF-24-1-0379. ILC acknowledges support in part by the NSF Center for Ultracold Atoms. This material is based upon work supported by the Department of Defense under Air Force Contract No. FA8702-15-D-0001. Any opinions, findings, conclusions, or recommendations expressed in this material are those of the author(s) and do not necessarily reflect the views of the Department of Defense.
        
\appendix*
\section{}
    \subsection{AC Stark Shift}
        We use the differential Stark shift, $\Delta_{d,s}$, between two levels, $\ket{s} = \ket{F_s, m_s}$ and $\ket{d} = \ket{F_d, m_d}$ in the S$_{1/2}$ and D$_{5/2}$ respectively, to determine the amplitude of the laser electric field.
        The AC Stark shift is calculated by \cite{Wineland_2003, manakov_1986}:
        \begin{align}
            \delta_i &= \frac{E^2_\ell}{4 \hbar^2} \sum_k \frac{\omega_{ik} \left| \bra{i} \bm{r} \cdot \hat{\epsilon}_\ell \ket{k} \right|}{\omega_{ik}^2 - \omega_\ell^2} \nonumber \\
            \Delta_{d,s} &= \delta_d - \delta_s
            \label{eq:diffacstark}
        \end{align}
        where the laser is described by the electric field amplitude $E_\ell$, frequency $\omega_\ell$, and polarization vector $\hat{\epsilon}_\ell$. The sum over $k$ includes all E1-allowed transitions, each with energy $\hbar \omega_{ik}$.
        
    \subsection{Polarization Measurements}
        We determine the polarization $\hat{\epsilon}_\ell$ and direction $\hat{k}$ of each laser beam by measuring the Rabi frequency of various transitions. For $674$ nm light, we use the quadrupole transitions in \ion{Sr}{88} between S$_{1/2}$ and D$_{5/2}$. For this single beam, we define $\hat{k} = (\sin{\phi}, 0, \cos{\phi})$, parameterized by its angle $\phi$ from the quantization axis provided by our magnetic field $\vec{B} \parallel (0,0,1)$, and $\hat{\epsilon}_\ell = (\cos{\gamma} \cos{\phi}, \sin{\gamma}, -\cos{\gamma}\sin{\phi})$ where $\gamma$ is the angle between $\hat{\epsilon}_\ell$ and $\vec{B}$ projected into the plane normal to $\hat{k}$. The quadrupole Rabi frequencies are calculated in terms of reduced matrix element $\bra{b} |r^2Q^{(2)}|\ket{a}$ between states $a$ and $b$:
         \begin{equation}
            \Omega^{(E2)}_{a,b} = \left|\frac{e E_\ell \omega_{ab}}{2\hbar c} \bra{b}|r^2Q^{(2)}|\ket{a}  \sum_{q=-2}^{2}
            \begin{Bmatrix}
             J_a & 2 & J_b \\
             -m_a & q & m_b
            \end{Bmatrix}_{(3)}
             c_{ij}^{(q)} \epsilon^i k^j \right|
        \label{eq:e2rabi}
        \end{equation}
        where $Q^{(2)}$ is the electric quadrupole operator, $c_{ij}^{(q)}$ is a second rank tensor (see A.6 \cite{James_1998}) and $c_{ij}^{(q)} \epsilon^i k^j$ its contraction, $\omega_{ab}$ is the transition frequency, and we have used a Wigner-3j symbol \cite{James_1998, Roos2000}. We can simplify this process for our needs by only considering the geometric factor as we only make use of the ratios of each $\Delta m_F$ process:
        \begin{align}
            \Omega_{\Delta_{mF}=0} &\propto \frac12 \left| \cos \gamma \sin{(2\phi)}\right| \\
            \Omega_{\Delta_{mF}=1} &\propto \frac{1}{\sqrt{6}} \left| \cos \gamma \cos{(2\phi)} - i \sin \gamma \cos \phi\right| \\
            \Omega_{\Delta_{mF}=2} &\propto \frac{1}{\sqrt{6}} \left| \frac12 \cos \gamma \sin{(2\phi)} - i \sin \gamma \sin \phi\right|
        \end{align}
        
        For the two other lasers at 461~nm and 617~nm we use Raman transitions in the S$_{1/2}$ and D$_{5/2}$ of \ion{Ba}{137} respectively. We define the second Raman laser beam with direction $\hat{k}' = (0, \sin{\phi'}, \cos{\phi'})$ at angle $\phi'$ to $\bm{B}$ and assume $\hat{k} \perp \hat{k}'$ and thus $\hat{\epsilon}_\ell' = (\sin{\gamma'}, \cos{\gamma'}\cos{\phi'}, -\cos{\gamma'}\sin{\phi'})$ with polarization projection $\gamma'$; $\hat{\epsilon}^*$ denotes the complex conjugate below. The Raman Rabi frequencies are \cite{moore_2023}:
        \begin{align}
            \Omega^{(Raman)}_{a,b} = \Biggl| \frac{e^2 E_\ell E_\ell'}{4\hbar^2} \sum_{k}  \biggl( &\frac{\bra{b}\bm{r}\cdot \hat{\epsilon}_\ell^*\ket{k}\bra{k}\bm{r}\cdot \hat{\epsilon}_\ell' \ket{a}}{\Delta_\Lambda} \nonumber \\ 
            + &\frac{\bra{b}\bm{r}\cdot \hat{\epsilon}_\ell \ket{k}\bra{k}\bm{r}\cdot \hat{\epsilon}_\ell'^* \ket{a}}{\Delta_V} \biggl) \Biggl|
        \label{eq:ramanrabi}
        \end{align}


    \subsection{Scattering rate for metastable states}
         We extend the ground state scattering model of Ozeri et al. \cite{ozeri_2005} by assuming a constant density of states (the scattered photon energy is constant), including only the $\Lambda$ scattering process, and only coupling to the nearest manifold (the metastable qubits we consider in $\mathrm{Ba}^+$ are housed in the $5D_{5/2}$ manifold and only couple to the $6P_{3/2}$ levels.).  This results in a scattering rate
        \begin{align}
            \Gamma_{i,f}^{Oz} &= \frac{E_\ell^2 \omega_{PD}^3}{12 \pi \varepsilon_0 \hbar^3 c^3} \sum_q \left| \sum_{k} \left(\frac{\langle f | \bm{r} \cdot \hat{\epsilon}_q | k \rangle \langle k | \bm{r} \cdot \hat{\epsilon}_\ell | i \rangle}{\Delta_\Lambda}\right) \right|^2, 
            \label{eq:ozscatter}
        \end{align}
        where $\omega_{PD}$ is the average transition frequency between the levels in the $P_{3/2}$ and $D_{5/2}$ manifolds.
        
\bibliography{refs}

\begin{thebibliography}{27}%
\makeatletter
\providecommand \@ifxundefined [1]{%
 \@ifx{#1\undefined}
}%
\providecommand \@ifnum [1]{%
 \ifnum #1\expandafter \@firstoftwo
 \else \expandafter \@secondoftwo
 \fi
}%
\providecommand \@ifx [1]{%
 \ifx #1\expandafter \@firstoftwo
 \else \expandafter \@secondoftwo
 \fi
}%
\providecommand \natexlab [1]{#1}%
\providecommand \enquote  [1]{``#1''}%
\providecommand \bibnamefont  [1]{#1}%
\providecommand \bibfnamefont [1]{#1}%
\providecommand \citenamefont [1]{#1}%
\providecommand \href@noop [0]{\@secondoftwo}%
\providecommand \href [0]{\begingroup \@sanitize@url \@href}%
\providecommand \@href[1]{\@@startlink{#1}\@@href}%
\providecommand \@@href[1]{\endgroup#1\@@endlink}%
\providecommand \@sanitize@url [0]{\catcode `\\12\catcode `\$12\catcode `\&12\catcode `\#12\catcode `\^12\catcode `\_12\catcode `\%12\relax}%
\providecommand \@@startlink[1]{}%
\providecommand \@@endlink[0]{}%
\providecommand \url  [0]{\begingroup\@sanitize@url \@url }%
\providecommand \@url [1]{\endgroup\@href {#1}{\urlprefix }}%
\providecommand \urlprefix  [0]{URL }%
\providecommand \Eprint [0]{\href }%
\providecommand \doibase [0]{https://doi.org/}%
\providecommand \selectlanguage [0]{\@gobble}%
\providecommand \bibinfo  [0]{\@secondoftwo}%
\providecommand \bibfield  [0]{\@secondoftwo}%
\providecommand \translation [1]{[#1]}%
\providecommand \BibitemOpen [0]{}%
\providecommand \bibitemStop [0]{}%
\providecommand \bibitemNoStop [0]{.\EOS\space}%
\providecommand \EOS [0]{\spacefactor3000\relax}%
\providecommand \BibitemShut  [1]{\csname bibitem#1\endcsname}%
\let\auto@bib@innerbib\@empty
\bibitem [{\citenamefont {Levine}\ \emph {et~al.}(2022)\citenamefont {Levine}, \citenamefont {Bluvstein}, \citenamefont {Keesling}, \citenamefont {Wang}, \citenamefont {Ebadi}, \citenamefont {Semeghini}, \citenamefont {Omran}, \citenamefont {Greiner}, \citenamefont {Vuleti\ifmmode~\acute{c}\else \'{c}\fi{}},\ and\ \citenamefont {Lukin}}]{Harry2022}%
  \BibitemOpen
  \bibfield  {author} {\bibinfo {author} {\bibfnamefont {H.}~\bibnamefont {Levine}}, \bibinfo {author} {\bibfnamefont {D.}~\bibnamefont {Bluvstein}}, \bibinfo {author} {\bibfnamefont {A.}~\bibnamefont {Keesling}}, \bibinfo {author} {\bibfnamefont {T.~T.}\ \bibnamefont {Wang}}, \bibinfo {author} {\bibfnamefont {S.}~\bibnamefont {Ebadi}}, \bibinfo {author} {\bibfnamefont {G.}~\bibnamefont {Semeghini}}, \bibinfo {author} {\bibfnamefont {A.}~\bibnamefont {Omran}}, \bibinfo {author} {\bibfnamefont {M.}~\bibnamefont {Greiner}}, \bibinfo {author} {\bibfnamefont {V.}~\bibnamefont {Vuleti\ifmmode~\acute{c}\else \'{c}\fi{}}},\ and\ \bibinfo {author} {\bibfnamefont {M.~D.}\ \bibnamefont {Lukin}},\ }\bibfield  {title} {\bibinfo {title} {Dispersive optical systems for scalable raman driving of hyperfine qubits},\ }\href {https://doi.org/10.1103/PhysRevA.105.032618} {\bibfield  {journal} {\bibinfo  {journal} {Phys. Rev. A}\ }\textbf {\bibinfo {volume} {105}},\ \bibinfo {pages} {032618} (\bibinfo {year} {2022})}\BibitemShut
  {NoStop}%
\bibitem [{\citenamefont {Ballance}\ \emph {et~al.}(2016)\citenamefont {Ballance}, \citenamefont {Harty}, \citenamefont {Linke}, \citenamefont {Sepiol},\ and\ \citenamefont {Lucas}}]{Ballance2016}%
  \BibitemOpen
  \bibfield  {author} {\bibinfo {author} {\bibfnamefont {C.~J.}\ \bibnamefont {Ballance}}, \bibinfo {author} {\bibfnamefont {T.~P.}\ \bibnamefont {Harty}}, \bibinfo {author} {\bibfnamefont {N.~M.}\ \bibnamefont {Linke}}, \bibinfo {author} {\bibfnamefont {M.~A.}\ \bibnamefont {Sepiol}},\ and\ \bibinfo {author} {\bibfnamefont {D.~M.}\ \bibnamefont {Lucas}},\ }\bibfield  {title} {\bibinfo {title} {High-fidelity quantum logic gates using trapped-ion hyperfine qubits},\ }\href {https://doi.org/10.1103/PhysRevLett.117.060504} {\bibfield  {journal} {\bibinfo  {journal} {Phys. Rev. Lett.}\ }\textbf {\bibinfo {volume} {117}},\ \bibinfo {pages} {060504} (\bibinfo {year} {2016})}\BibitemShut {NoStop}%
\bibitem [{\citenamefont {Gaebler}\ \emph {et~al.}(2016)\citenamefont {Gaebler}, \citenamefont {Tan}, \citenamefont {Lin}, \citenamefont {Wan}, \citenamefont {Bowler}, \citenamefont {Keith}, \citenamefont {Glancy}, \citenamefont {Coakley}, \citenamefont {Knill}, \citenamefont {Leibfried},\ and\ \citenamefont {Wineland}}]{gaebler_high-fidelity_2016}%
  \BibitemOpen
  \bibfield  {author} {\bibinfo {author} {\bibfnamefont {J.~P.}\ \bibnamefont {Gaebler}}, \bibinfo {author} {\bibfnamefont {T.~R.}\ \bibnamefont {Tan}}, \bibinfo {author} {\bibfnamefont {Y.}~\bibnamefont {Lin}}, \bibinfo {author} {\bibfnamefont {Y.}~\bibnamefont {Wan}}, \bibinfo {author} {\bibfnamefont {R.}~\bibnamefont {Bowler}}, \bibinfo {author} {\bibfnamefont {A.~C.}\ \bibnamefont {Keith}}, \bibinfo {author} {\bibfnamefont {S.}~\bibnamefont {Glancy}}, \bibinfo {author} {\bibfnamefont {K.}~\bibnamefont {Coakley}}, \bibinfo {author} {\bibfnamefont {E.}~\bibnamefont {Knill}}, \bibinfo {author} {\bibfnamefont {D.}~\bibnamefont {Leibfried}},\ and\ \bibinfo {author} {\bibfnamefont {D.~J.}\ \bibnamefont {Wineland}},\ }\bibfield  {title} {\bibinfo {title} {High-fidelity universal gate set for ${^{9}\mathrm{Be}}^{+}$ ion qubits},\ }\href {https://doi.org/10.1103/PhysRevLett.117.060505} {\bibfield  {journal} {\bibinfo  {journal} {Phys. Rev. Lett.}\ }\textbf {\bibinfo {volume} {117}},\ \bibinfo {pages} {060505}
  (\bibinfo {year} {2016})}\BibitemShut {NoStop}%
\bibitem [{\citenamefont {Kerman}\ \emph {et~al.}(2000)\citenamefont {Kerman}, \citenamefont {Vuleti\ifmmode~\acute{c}\else \'{c}\fi{}}, \citenamefont {Chin},\ and\ \citenamefont {Chu}}]{Kerman2000}%
  \BibitemOpen
  \bibfield  {author} {\bibinfo {author} {\bibfnamefont {A.~J.}\ \bibnamefont {Kerman}}, \bibinfo {author} {\bibfnamefont {V.}~\bibnamefont {Vuleti\ifmmode~\acute{c}\else \'{c}\fi{}}}, \bibinfo {author} {\bibfnamefont {C.}~\bibnamefont {Chin}},\ and\ \bibinfo {author} {\bibfnamefont {S.}~\bibnamefont {Chu}},\ }\bibfield  {title} {\bibinfo {title} {Beyond optical molasses: 3d raman sideband cooling of atomic cesium to high phase-space density},\ }\href {https://doi.org/10.1103/PhysRevLett.84.439} {\bibfield  {journal} {\bibinfo  {journal} {Phys. Rev. Lett.}\ }\textbf {\bibinfo {volume} {84}},\ \bibinfo {pages} {439} (\bibinfo {year} {2000})}\BibitemShut {NoStop}%
\bibitem [{\citenamefont {Berto}\ \emph {et~al.}(2014)\citenamefont {Berto}, \citenamefont {Andresen},\ and\ \citenamefont {Rigneault}}]{Berto2014}%
  \BibitemOpen
  \bibfield  {author} {\bibinfo {author} {\bibfnamefont {P.}~\bibnamefont {Berto}}, \bibinfo {author} {\bibfnamefont {E.~R.}\ \bibnamefont {Andresen}},\ and\ \bibinfo {author} {\bibfnamefont {H.}~\bibnamefont {Rigneault}},\ }\bibfield  {title} {\bibinfo {title} {Background-free stimulated raman spectroscopy and microscopy},\ }\href {https://doi.org/10.1103/PhysRevLett.112.053905} {\bibfield  {journal} {\bibinfo  {journal} {Phys. Rev. Lett.}\ }\textbf {\bibinfo {volume} {112}},\ \bibinfo {pages} {053905} (\bibinfo {year} {2014})}\BibitemShut {NoStop}%
\bibitem [{\citenamefont {Allcock}\ \emph {et~al.}(2021)\citenamefont {Allcock}, \citenamefont {Campbell}, \citenamefont {Chiaverini}, \citenamefont {Chuang}, \citenamefont {Hudson}, \citenamefont {Moore}, \citenamefont {Ransford}, \citenamefont {Roman}, \citenamefont {Sage},\ and\ \citenamefont {Wineland}}]{allcock_omg_2021}%
  \BibitemOpen
  \bibfield  {author} {\bibinfo {author} {\bibfnamefont {D.~T.~C.}\ \bibnamefont {Allcock}}, \bibinfo {author} {\bibfnamefont {W.~C.}\ \bibnamefont {Campbell}}, \bibinfo {author} {\bibfnamefont {J.}~\bibnamefont {Chiaverini}}, \bibinfo {author} {\bibfnamefont {I.~L.}\ \bibnamefont {Chuang}}, \bibinfo {author} {\bibfnamefont {E.~R.}\ \bibnamefont {Hudson}}, \bibinfo {author} {\bibfnamefont {I.~D.}\ \bibnamefont {Moore}}, \bibinfo {author} {\bibfnamefont {A.}~\bibnamefont {Ransford}}, \bibinfo {author} {\bibfnamefont {C.}~\bibnamefont {Roman}}, \bibinfo {author} {\bibfnamefont {J.~M.}\ \bibnamefont {Sage}},\ and\ \bibinfo {author} {\bibfnamefont {D.~J.}\ \bibnamefont {Wineland}},\ }\bibfield  {title} {\bibinfo {title} {\textit{omg} blueprint for trapped ion quantum computing with metastable states},\ }\href {https://doi.org/10.1063/5.0069544} {\bibfield  {journal} {\bibinfo  {journal} {Applied Physics Letters}\ }\textbf {\bibinfo {volume} {119}},\ \bibinfo {pages} {214002} (\bibinfo {year} {2021})}\BibitemShut
  {NoStop}%
\bibitem [{\citenamefont {Yang}\ \emph {et~al.}(2022)\citenamefont {Yang}, \citenamefont {Ma}, \citenamefont {Wu}, \citenamefont {Wang}, \citenamefont {Cao}, \citenamefont {Guo}, \citenamefont {Huang}, \citenamefont {Feng}, \citenamefont {Zhou},\ and\ \citenamefont {Duan}}]{yang_realizing_2022}%
  \BibitemOpen
  \bibfield  {author} {\bibinfo {author} {\bibfnamefont {H.-X.}\ \bibnamefont {Yang}}, \bibinfo {author} {\bibfnamefont {J.-Y.}\ \bibnamefont {Ma}}, \bibinfo {author} {\bibfnamefont {Y.-K.}\ \bibnamefont {Wu}}, \bibinfo {author} {\bibfnamefont {Y.}~\bibnamefont {Wang}}, \bibinfo {author} {\bibfnamefont {M.-M.}\ \bibnamefont {Cao}}, \bibinfo {author} {\bibfnamefont {W.-X.}\ \bibnamefont {Guo}}, \bibinfo {author} {\bibfnamefont {Y.-Y.}\ \bibnamefont {Huang}}, \bibinfo {author} {\bibfnamefont {L.}~\bibnamefont {Feng}}, \bibinfo {author} {\bibfnamefont {Z.-C.}\ \bibnamefont {Zhou}},\ and\ \bibinfo {author} {\bibfnamefont {L.-M.}\ \bibnamefont {Duan}},\ }\bibfield  {title} {\bibinfo {title} {Realizing coherently convertible dual-type qubits with the same ion species},\ }\href {https://doi.org/10.1038/s41567-022-01661-5} {\bibfield  {journal} {\bibinfo  {journal} {Nature Physics}\ }\textbf {\bibinfo {volume} {18}},\ \bibinfo {pages} {1058} (\bibinfo {year} {2022})}\BibitemShut {NoStop}%
\bibitem [{\citenamefont {DeBry}\ \emph {et~al.}(2025)\citenamefont {DeBry}, \citenamefont {Meister}, \citenamefont {Martinez}, \citenamefont {Bruzewicz}, \citenamefont {Shi}, \citenamefont {Reens}, \citenamefont {McConnell}, \citenamefont {Chuang},\ and\ \citenamefont {Chiaverini}}]{debry2025errorcorrectionlogicalqubit}%
  \BibitemOpen
  \bibfield  {author} {\bibinfo {author} {\bibfnamefont {K.}~\bibnamefont {DeBry}}, \bibinfo {author} {\bibfnamefont {N.}~\bibnamefont {Meister}}, \bibinfo {author} {\bibfnamefont {A.~V.}\ \bibnamefont {Martinez}}, \bibinfo {author} {\bibfnamefont {C.~D.}\ \bibnamefont {Bruzewicz}}, \bibinfo {author} {\bibfnamefont {X.}~\bibnamefont {Shi}}, \bibinfo {author} {\bibfnamefont {D.}~\bibnamefont {Reens}}, \bibinfo {author} {\bibfnamefont {R.}~\bibnamefont {McConnell}}, \bibinfo {author} {\bibfnamefont {I.~L.}\ \bibnamefont {Chuang}},\ and\ \bibinfo {author} {\bibfnamefont {J.}~\bibnamefont {Chiaverini}},\ }\href {https://arxiv.org/abs/2503.13908} {\bibinfo {title} {Error correction of a logical qubit encoded in a single atomic ion}} (\bibinfo {year} {2025}),\ \Eprint {https://arxiv.org/abs/2503.13908} {arXiv:2503.13908 [quant-ph]} \BibitemShut {NoStop}%
\bibitem [{\citenamefont {Quinn}\ \emph {et~al.}(2024)\citenamefont {Quinn}, \citenamefont {Gregory}, \citenamefont {Moore}, \citenamefont {Brudney}, \citenamefont {Metzner}, \citenamefont {Ritchie}, \citenamefont {O'Reilly}, \citenamefont {Wineland},\ and\ \citenamefont {Allcock}}]{quinn2024highfidelityentanglementmetastabletrappedion}%
  \BibitemOpen
  \bibfield  {author} {\bibinfo {author} {\bibfnamefont {A.}~\bibnamefont {Quinn}}, \bibinfo {author} {\bibfnamefont {G.~J.}\ \bibnamefont {Gregory}}, \bibinfo {author} {\bibfnamefont {I.~D.}\ \bibnamefont {Moore}}, \bibinfo {author} {\bibfnamefont {S.}~\bibnamefont {Brudney}}, \bibinfo {author} {\bibfnamefont {J.}~\bibnamefont {Metzner}}, \bibinfo {author} {\bibfnamefont {E.~R.}\ \bibnamefont {Ritchie}}, \bibinfo {author} {\bibfnamefont {J.}~\bibnamefont {O'Reilly}}, \bibinfo {author} {\bibfnamefont {D.~J.}\ \bibnamefont {Wineland}},\ and\ \bibinfo {author} {\bibfnamefont {D.~T.~C.}\ \bibnamefont {Allcock}},\ }\href {https://arxiv.org/abs/2411.12727} {\bibinfo {title} {High-fidelity entanglement of metastable trapped-ion qubits with integrated erasure conversion}} (\bibinfo {year} {2024}),\ \Eprint {https://arxiv.org/abs/2411.12727} {arXiv:2411.12727 [physics.atom-ph]} \BibitemShut {NoStop}%
\bibitem [{\citenamefont {DeBry}\ \emph {et~al.}(2023)\citenamefont {DeBry}, \citenamefont {Sinanan-Singh}, \citenamefont {Bruzewicz}, \citenamefont {Reens}, \citenamefont {Kim}, \citenamefont {Roychowdhury}, \citenamefont {McConnell}, \citenamefont {Chuang},\ and\ \citenamefont {Chiaverini}}]{debry_experimental_2023}%
  \BibitemOpen
  \bibfield  {author} {\bibinfo {author} {\bibfnamefont {K.}~\bibnamefont {DeBry}}, \bibinfo {author} {\bibfnamefont {J.}~\bibnamefont {Sinanan-Singh}}, \bibinfo {author} {\bibfnamefont {C.~D.}\ \bibnamefont {Bruzewicz}}, \bibinfo {author} {\bibfnamefont {D.}~\bibnamefont {Reens}}, \bibinfo {author} {\bibfnamefont {M.~E.}\ \bibnamefont {Kim}}, \bibinfo {author} {\bibfnamefont {M.~P.}\ \bibnamefont {Roychowdhury}}, \bibinfo {author} {\bibfnamefont {R.}~\bibnamefont {McConnell}}, \bibinfo {author} {\bibfnamefont {I.~L.}\ \bibnamefont {Chuang}},\ and\ \bibinfo {author} {\bibfnamefont {J.}~\bibnamefont {Chiaverini}},\ }\bibfield  {title} {\bibinfo {title} {Experimental quantum channel discrimination using metastable states of a trapped ion},\ }\href {https://doi.org/10.1103/PhysRevLett.131.170602} {\bibfield  {journal} {\bibinfo  {journal} {Physical Review Letters}\ }\textbf {\bibinfo {volume} {131}},\ \bibinfo {pages} {170602} (\bibinfo {year} {2023})}\BibitemShut {NoStop}%
\bibitem [{\citenamefont {Vizvary}\ \emph {et~al.}(2024)\citenamefont {Vizvary}, \citenamefont {Wall}, \citenamefont {Boguslawski}, \citenamefont {Bareian}, \citenamefont {Derevianko}, \citenamefont {Campbell},\ and\ \citenamefont {Hudson}}]{vizvary_eliminating_2024}%
  \BibitemOpen
  \bibfield  {author} {\bibinfo {author} {\bibfnamefont {S.~R.}\ \bibnamefont {Vizvary}}, \bibinfo {author} {\bibfnamefont {Z.~J.}\ \bibnamefont {Wall}}, \bibinfo {author} {\bibfnamefont {M.~J.}\ \bibnamefont {Boguslawski}}, \bibinfo {author} {\bibfnamefont {M.}~\bibnamefont {Bareian}}, \bibinfo {author} {\bibfnamefont {A.}~\bibnamefont {Derevianko}}, \bibinfo {author} {\bibfnamefont {W.~C.}\ \bibnamefont {Campbell}},\ and\ \bibinfo {author} {\bibfnamefont {E.~R.}\ \bibnamefont {Hudson}},\ }\bibfield  {title} {\bibinfo {title} {Eliminating qubit-type cross-talk in the $omg$ protocol},\ }\href {https://doi.org/10.1103/PhysRevLett.132.263201} {\bibfield  {journal} {\bibinfo  {journal} {Phys. Rev. Lett.}\ }\textbf {\bibinfo {volume} {132}},\ \bibinfo {pages} {263201} (\bibinfo {year} {2024})}\BibitemShut {NoStop}%
\bibitem [{\citenamefont {Sotirova}\ \emph {et~al.}(2024)\citenamefont {Sotirova}, \citenamefont {Leppard}, \citenamefont {Vazquez-Brennan}, \citenamefont {Decoppet}, \citenamefont {Pokorny}, \citenamefont {Malinowski},\ and\ \citenamefont {Ballance}}]{sotirova2024highfidelityheraldedquantumstate}%
  \BibitemOpen
  \bibfield  {author} {\bibinfo {author} {\bibfnamefont {A.~S.}\ \bibnamefont {Sotirova}}, \bibinfo {author} {\bibfnamefont {J.~D.}\ \bibnamefont {Leppard}}, \bibinfo {author} {\bibfnamefont {A.}~\bibnamefont {Vazquez-Brennan}}, \bibinfo {author} {\bibfnamefont {S.~M.}\ \bibnamefont {Decoppet}}, \bibinfo {author} {\bibfnamefont {F.}~\bibnamefont {Pokorny}}, \bibinfo {author} {\bibfnamefont {M.}~\bibnamefont {Malinowski}},\ and\ \bibinfo {author} {\bibfnamefont {C.~J.}\ \bibnamefont {Ballance}},\ }\href {https://arxiv.org/abs/2409.05805} {\bibinfo {title} {High-fidelity heralded quantum state preparation and measurement}} (\bibinfo {year} {2024}),\ \Eprint {https://arxiv.org/abs/2409.05805} {arXiv:2409.05805 [quant-ph]} \BibitemShut {NoStop}%
\bibitem [{\citenamefont {B\ifmmode \u{a}\else \u{a}\fi{}z\ifmmode~\u{a}\else \u{a}\fi{}van}\ \emph {et~al.}(2023)\citenamefont {B\ifmmode \u{a}\else \u{a}\fi{}z\ifmmode~\u{a}\else \u{a}\fi{}van}, \citenamefont {Saner}, \citenamefont {Minder}, \citenamefont {Hughes}, \citenamefont {Sutherland}, \citenamefont {Lucas}, \citenamefont {Srinivas},\ and\ \citenamefont {Ballance}}]{bazavan_synthesizing_2023}%
  \BibitemOpen
  \bibfield  {author} {\bibinfo {author} {\bibfnamefont {O.}~\bibnamefont {B\ifmmode \u{a}\else \u{a}\fi{}z\ifmmode~\u{a}\else \u{a}\fi{}van}}, \bibinfo {author} {\bibfnamefont {S.}~\bibnamefont {Saner}}, \bibinfo {author} {\bibfnamefont {M.}~\bibnamefont {Minder}}, \bibinfo {author} {\bibfnamefont {A.~C.}\ \bibnamefont {Hughes}}, \bibinfo {author} {\bibfnamefont {R.~T.}\ \bibnamefont {Sutherland}}, \bibinfo {author} {\bibfnamefont {D.~M.}\ \bibnamefont {Lucas}}, \bibinfo {author} {\bibfnamefont {R.}~\bibnamefont {Srinivas}},\ and\ \bibinfo {author} {\bibfnamefont {C.~J.}\ \bibnamefont {Ballance}},\ }\bibfield  {title} {\bibinfo {title} {Synthesizing a ${\stackrel{\ifmmode \hat{}\else \^{}\fi{}}{\ensuremath{\sigma}}}_{z}$ spin-dependent force for optical, metastable, and ground-state trapped-ion qubits},\ }\href {https://doi.org/10.1103/PhysRevA.107.022617} {\bibfield  {journal} {\bibinfo  {journal} {Phys. Rev. A}\ }\textbf {\bibinfo {volume} {107}},\ \bibinfo {pages} {022617} (\bibinfo {year}
  {2023})}\BibitemShut {NoStop}%
\bibitem [{\citenamefont {Wang}\ \emph {et~al.}(2025)\citenamefont {Wang}, \citenamefont {Huang}, \citenamefont {Zhang}, \citenamefont {Hu}, \citenamefont {Mao}, \citenamefont {Hou}, \citenamefont {Wu}, \citenamefont {Zhou},\ and\ \citenamefont {Duan}}]{wang_experimental_2025}%
  \BibitemOpen
  \bibfield  {author} {\bibinfo {author} {\bibfnamefont {C.}~\bibnamefont {Wang}}, \bibinfo {author} {\bibfnamefont {C.}~\bibnamefont {Huang}}, \bibinfo {author} {\bibfnamefont {H.}~\bibnamefont {Zhang}}, \bibinfo {author} {\bibfnamefont {H.}~\bibnamefont {Hu}}, \bibinfo {author} {\bibfnamefont {Z.}~\bibnamefont {Mao}}, \bibinfo {author} {\bibfnamefont {P.}~\bibnamefont {Hou}}, \bibinfo {author} {\bibfnamefont {Y.}~\bibnamefont {Wu}}, \bibinfo {author} {\bibfnamefont {Z.}~\bibnamefont {Zhou}},\ and\ \bibinfo {author} {\bibfnamefont {L.}~\bibnamefont {Duan}},\ }\bibfield  {title} {\bibinfo {title} {Experimental realization of direct entangling gates between dual-type qubits},\ }\href {https://doi.org/10.1103/PhysRevLett.134.010601} {\bibfield  {journal} {\bibinfo  {journal} {Phys. Rev. Lett.}\ }\textbf {\bibinfo {volume} {134}},\ \bibinfo {pages} {010601} (\bibinfo {year} {2025})}\BibitemShut {NoStop}%
\bibitem [{\citenamefont {Ozeri}\ \emph {et~al.}(2005)\citenamefont {Ozeri}, \citenamefont {Langer}, \citenamefont {Jost}, \citenamefont {DeMarco}, \citenamefont {Ben-Kish}, \citenamefont {Blakestad}, \citenamefont {Britton}, \citenamefont {Chiaverini}, \citenamefont {Itano}, \citenamefont {Hume}, \citenamefont {Leibfried}, \citenamefont {Rosenband}, \citenamefont {Schmidt},\ and\ \citenamefont {Wineland}}]{ozeri_2005}%
  \BibitemOpen
  \bibfield  {author} {\bibinfo {author} {\bibfnamefont {R.}~\bibnamefont {Ozeri}}, \bibinfo {author} {\bibfnamefont {C.}~\bibnamefont {Langer}}, \bibinfo {author} {\bibfnamefont {J.~D.}\ \bibnamefont {Jost}}, \bibinfo {author} {\bibfnamefont {B.}~\bibnamefont {DeMarco}}, \bibinfo {author} {\bibfnamefont {A.}~\bibnamefont {Ben-Kish}}, \bibinfo {author} {\bibfnamefont {B.~R.}\ \bibnamefont {Blakestad}}, \bibinfo {author} {\bibfnamefont {J.}~\bibnamefont {Britton}}, \bibinfo {author} {\bibfnamefont {J.}~\bibnamefont {Chiaverini}}, \bibinfo {author} {\bibfnamefont {W.~M.}\ \bibnamefont {Itano}}, \bibinfo {author} {\bibfnamefont {D.~B.}\ \bibnamefont {Hume}}, \bibinfo {author} {\bibfnamefont {D.}~\bibnamefont {Leibfried}}, \bibinfo {author} {\bibfnamefont {T.}~\bibnamefont {Rosenband}}, \bibinfo {author} {\bibfnamefont {P.~O.}\ \bibnamefont {Schmidt}},\ and\ \bibinfo {author} {\bibfnamefont {D.~J.}\ \bibnamefont {Wineland}},\ }\bibfield  {title} {\bibinfo {title} {Hyperfine coherence in the presence of
  spontaneous photon scattering},\ }\href {https://doi.org/10.1103/PhysRevLett.95.030403} {\bibfield  {journal} {\bibinfo  {journal} {Phys. Rev. Lett.}\ }\textbf {\bibinfo {volume} {95}},\ \bibinfo {pages} {030403} (\bibinfo {year} {2005})}\BibitemShut {NoStop}%
\bibitem [{\citenamefont {Moore}\ \emph {et~al.}(2023)\citenamefont {Moore}, \citenamefont {Campbell}, \citenamefont {Hudson}, \citenamefont {Boguslawski}, \citenamefont {Wineland},\ and\ \citenamefont {Allcock}}]{moore_2023}%
  \BibitemOpen
  \bibfield  {author} {\bibinfo {author} {\bibfnamefont {I.~D.}\ \bibnamefont {Moore}}, \bibinfo {author} {\bibfnamefont {W.~C.}\ \bibnamefont {Campbell}}, \bibinfo {author} {\bibfnamefont {E.~R.}\ \bibnamefont {Hudson}}, \bibinfo {author} {\bibfnamefont {M.~J.}\ \bibnamefont {Boguslawski}}, \bibinfo {author} {\bibfnamefont {D.~J.}\ \bibnamefont {Wineland}},\ and\ \bibinfo {author} {\bibfnamefont {D.~T.~C.}\ \bibnamefont {Allcock}},\ }\bibfield  {title} {\bibinfo {title} {Photon scattering errors during stimulated raman transitions in trapped-ion qubits},\ }\href {https://doi.org/10.1103/PhysRevA.107.032413} {\bibfield  {journal} {\bibinfo  {journal} {Phys. Rev. A}\ }\textbf {\bibinfo {volume} {107}},\ \bibinfo {pages} {032413} (\bibinfo {year} {2023})}\BibitemShut {NoStop}%
\bibitem [{\citenamefont {Boguslawski}\ \emph {et~al.}(2023)\citenamefont {Boguslawski}, \citenamefont {Wall}, \citenamefont {Vizvary}, \citenamefont {Moore}, \citenamefont {Bareian}, \citenamefont {Allcock}, \citenamefont {Wineland}, \citenamefont {Hudson},\ and\ \citenamefont {Campbell}}]{boguslawski_2023}%
  \BibitemOpen
  \bibfield  {author} {\bibinfo {author} {\bibfnamefont {M.~J.}\ \bibnamefont {Boguslawski}}, \bibinfo {author} {\bibfnamefont {Z.~J.}\ \bibnamefont {Wall}}, \bibinfo {author} {\bibfnamefont {S.~R.}\ \bibnamefont {Vizvary}}, \bibinfo {author} {\bibfnamefont {I.~D.}\ \bibnamefont {Moore}}, \bibinfo {author} {\bibfnamefont {M.}~\bibnamefont {Bareian}}, \bibinfo {author} {\bibfnamefont {D.~T.~C.}\ \bibnamefont {Allcock}}, \bibinfo {author} {\bibfnamefont {D.~J.}\ \bibnamefont {Wineland}}, \bibinfo {author} {\bibfnamefont {E.~R.}\ \bibnamefont {Hudson}},\ and\ \bibinfo {author} {\bibfnamefont {W.~C.}\ \bibnamefont {Campbell}},\ }\bibfield  {title} {\bibinfo {title} {Raman scattering errors in stimulated-raman-induced logic gates in $^{133}{\mathrm{ba}}^{+}$},\ }\href {https://doi.org/10.1103/PhysRevLett.131.063001} {\bibfield  {journal} {\bibinfo  {journal} {Phys. Rev. Lett.}\ }\textbf {\bibinfo {volume} {131}},\ \bibinfo {pages} {063001} (\bibinfo {year} {2023})}\BibitemShut {NoStop}%
\bibitem [{\citenamefont {Barakhshan}\ \emph {et~al.}()\citenamefont {Barakhshan}, \citenamefont {Marrs}, \citenamefont {Bhosale}, \citenamefont {Arora}, \citenamefont {Eigenmann},\ and\ \citenamefont {Safronova}}]{UDportal}%
  \BibitemOpen
  \bibfield  {author} {\bibinfo {author} {\bibfnamefont {P.}~\bibnamefont {Barakhshan}}, \bibinfo {author} {\bibfnamefont {A.}~\bibnamefont {Marrs}}, \bibinfo {author} {\bibfnamefont {A.}~\bibnamefont {Bhosale}}, \bibinfo {author} {\bibfnamefont {B.}~\bibnamefont {Arora}}, \bibinfo {author} {\bibfnamefont {R.}~\bibnamefont {Eigenmann}},\ and\ \bibinfo {author} {\bibfnamefont {M.~S.}\ \bibnamefont {Safronova}},\ }\href@noop {} {}\bibinfo {howpublished} {{\textit{Portal for High-Precision Atomic Data and Computation}} (version 2.0). University of Delaware, Newark, DE, USA. URL: {https://www.udel.edu/atom} [February 2022].}\BibitemShut {Stop}%
\bibitem [{\citenamefont {Kramida}\ \emph {et~al.}(2024)\citenamefont {Kramida}, \citenamefont {{Yu.~Ralchenko}}, \citenamefont {Reader},\ and\ \citenamefont {{and NIST ASD Team}}}]{NIST_ASD}%
  \BibitemOpen
  \bibfield  {author} {\bibinfo {author} {\bibfnamefont {A.}~\bibnamefont {Kramida}}, \bibinfo {author} {\bibnamefont {{Yu.~Ralchenko}}}, \bibinfo {author} {\bibfnamefont {J.}~\bibnamefont {Reader}},\ and\ \bibinfo {author} {\bibnamefont {{and NIST ASD Team}}},\ }\href@noop {} {}\bibinfo {howpublished} {{NIST Atomic Spectra Database (ver. 5.12), [Online]. Available: {\tt{https://physics.nist.gov/asd}} [2025, April 15]. National Institute of Standards and Technology, Gaithersburg, MD.}} (\bibinfo {year} {2024})\BibitemShut {NoStop}%
\bibitem [{\citenamefont {Shi}\ \emph {et~al.}(2025)\citenamefont {Shi}, \citenamefont {Sinanan-Singh}, \citenamefont {DeBry}, \citenamefont {Todaro}, \citenamefont {Chuang},\ and\ \citenamefont {Chiaverini}}]{shi_long-lived_2025}%
  \BibitemOpen
  \bibfield  {author} {\bibinfo {author} {\bibfnamefont {X.}~\bibnamefont {Shi}}, \bibinfo {author} {\bibfnamefont {J.}~\bibnamefont {Sinanan-Singh}}, \bibinfo {author} {\bibfnamefont {K.}~\bibnamefont {DeBry}}, \bibinfo {author} {\bibfnamefont {S.~L.}\ \bibnamefont {Todaro}}, \bibinfo {author} {\bibfnamefont {I.~L.}\ \bibnamefont {Chuang}},\ and\ \bibinfo {author} {\bibfnamefont {J.}~\bibnamefont {Chiaverini}},\ }\bibfield  {title} {\bibinfo {title} {Long-lived metastable-qubit memory},\ }\href {https://doi.org/10.1103/PhysRevA.111.L020601} {\bibfield  {journal} {\bibinfo  {journal} {Phys. Rev. A}\ }\textbf {\bibinfo {volume} {111}},\ \bibinfo {pages} {L020601} (\bibinfo {year} {2025})}\BibitemShut {NoStop}%
\bibitem [{\citenamefont {An}\ \emph {et~al.}(2022)\citenamefont {An}, \citenamefont {Ransford}, \citenamefont {Schaffer}, \citenamefont {Sletten}, \citenamefont {Gaebler}, \citenamefont {Hostetter},\ and\ \citenamefont {Vittorini}}]{an_high_2022}%
  \BibitemOpen
  \bibfield  {author} {\bibinfo {author} {\bibfnamefont {F.~A.}\ \bibnamefont {An}}, \bibinfo {author} {\bibfnamefont {A.}~\bibnamefont {Ransford}}, \bibinfo {author} {\bibfnamefont {A.}~\bibnamefont {Schaffer}}, \bibinfo {author} {\bibfnamefont {L.~R.}\ \bibnamefont {Sletten}}, \bibinfo {author} {\bibfnamefont {J.}~\bibnamefont {Gaebler}}, \bibinfo {author} {\bibfnamefont {J.}~\bibnamefont {Hostetter}},\ and\ \bibinfo {author} {\bibfnamefont {G.}~\bibnamefont {Vittorini}},\ }\bibfield  {title} {\bibinfo {title} {High {Fidelity} {State} {Preparation} and {Measurement} of {Ion} {Hyperfine} {Qubits} with {I} {\textgreater} 1 2},\ }\href {https://doi.org/10.1103/PhysRevLett.129.130501} {\bibfield  {journal} {\bibinfo  {journal} {Physical Review Letters}\ }\textbf {\bibinfo {volume} {129}},\ \bibinfo {pages} {130501} (\bibinfo {year} {2022})}\BibitemShut {NoStop}%
\bibitem [{\citenamefont {Roos}(2000)}]{Roos2000}%
  \BibitemOpen
  \bibfield  {author} {\bibinfo {author} {\bibfnamefont {C.~F.}\ \bibnamefont {Roos}},\ }\emph {\bibinfo {title} {Controlling the quantum stateof trapped ions}},\ \href {https://www.quantumoptics.at/images/publications/dissertation/roos-diss.pdf} {Ph.D. thesis},\ \bibinfo  {school} {University of Innsbruck} (\bibinfo {year} {2000})\BibitemShut {NoStop}%
\bibitem [{\citenamefont {James}(1998)}]{James_1998}%
  \BibitemOpen
  \bibfield  {author} {\bibinfo {author} {\bibfnamefont {D.~F.~V.}\ \bibnamefont {James}},\ }\bibfield  {title} {\bibinfo {title} {Quantum dynamics of cold trapped ions with application to quantum computation},\ }\href {https://doi.org/10.1007/s003400050373} {\bibfield  {journal} {\bibinfo  {journal} {Applied Physics B: Lasers and Optics}\ }\textbf {\bibinfo {volume} {66}},\ \bibinfo {pages} {181–190} (\bibinfo {year} {1998})}\BibitemShut {NoStop}%
\bibitem [{\citenamefont {S\o{}rensen}\ and\ \citenamefont {M\o{}lmer}(2000)}]{MS_gate}%
  \BibitemOpen
  \bibfield  {author} {\bibinfo {author} {\bibfnamefont {A.}~\bibnamefont {S\o{}rensen}}\ and\ \bibinfo {author} {\bibfnamefont {K.}~\bibnamefont {M\o{}lmer}},\ }\bibfield  {title} {\bibinfo {title} {Entanglement and quantum computation with ions in thermal motion},\ }\href {https://doi.org/10.1103/PhysRevA.62.022311} {\bibfield  {journal} {\bibinfo  {journal} {Phys. Rev. A}\ }\textbf {\bibinfo {volume} {62}},\ \bibinfo {pages} {022311} (\bibinfo {year} {2000})}\BibitemShut {NoStop}%
\bibitem [{\citenamefont {Moore}\ \emph {et~al.}(2025)\citenamefont {Moore}, \citenamefont {Quinn}, \citenamefont {O’Reilly}, \citenamefont {Metzner}, \citenamefont {Brudney}, \citenamefont {Gregory}, \citenamefont {Wineland},\ and\ \citenamefont {Allcock}}]{moore_spontaneous_2025}%
  \BibitemOpen
  \bibfield  {author} {\bibinfo {author} {\bibfnamefont {I.~D.}\ \bibnamefont {Moore}}, \bibinfo {author} {\bibfnamefont {A.}~\bibnamefont {Quinn}}, \bibinfo {author} {\bibfnamefont {J.}~\bibnamefont {O’Reilly}}, \bibinfo {author} {\bibfnamefont {J.}~\bibnamefont {Metzner}}, \bibinfo {author} {\bibfnamefont {S.}~\bibnamefont {Brudney}}, \bibinfo {author} {\bibfnamefont {G.~J.}\ \bibnamefont {Gregory}}, \bibinfo {author} {\bibfnamefont {D.~J.}\ \bibnamefont {Wineland}},\ and\ \bibinfo {author} {\bibfnamefont {D.~T.~C.}\ \bibnamefont {Allcock}},\ }\href {https://arxiv.org/abs/2505.04854} {\bibinfo {title} {Spontaneous raman scattering out of a metastable atomic qubit}} (\bibinfo {year} {2025}),\ \Eprint {https://arxiv.org/abs/2505.04854} {arXiv:2505.04854 [quant-ph]} \BibitemShut {NoStop}%
\bibitem [{\citenamefont {Wineland}\ \emph {et~al.}(2003)\citenamefont {Wineland}, \citenamefont {Barrett}, \citenamefont {Britton}, \citenamefont {Chiaverini}, \citenamefont {DeMarco}, \citenamefont {Itano}, \citenamefont {Jelenković}, \citenamefont {Langer}, \citenamefont {Leibfried}, \citenamefont {Meyer}, \citenamefont {Rosenband},\ and\ \citenamefont {Schätz}}]{Wineland_2003}%
  \BibitemOpen
  \bibfield  {author} {\bibinfo {author} {\bibfnamefont {D.~J.}\ \bibnamefont {Wineland}}, \bibinfo {author} {\bibfnamefont {M.}~\bibnamefont {Barrett}}, \bibinfo {author} {\bibfnamefont {J.}~\bibnamefont {Britton}}, \bibinfo {author} {\bibfnamefont {J.}~\bibnamefont {Chiaverini}}, \bibinfo {author} {\bibfnamefont {B.}~\bibnamefont {DeMarco}}, \bibinfo {author} {\bibfnamefont {W.~M.}\ \bibnamefont {Itano}}, \bibinfo {author} {\bibfnamefont {B.}~\bibnamefont {Jelenković}}, \bibinfo {author} {\bibfnamefont {C.}~\bibnamefont {Langer}}, \bibinfo {author} {\bibfnamefont {D.}~\bibnamefont {Leibfried}}, \bibinfo {author} {\bibfnamefont {V.}~\bibnamefont {Meyer}}, \bibinfo {author} {\bibfnamefont {T.}~\bibnamefont {Rosenband}},\ and\ \bibinfo {author} {\bibfnamefont {T.}~\bibnamefont {Schätz}},\ }\bibfield  {title} {\bibinfo {title} {Quantum information processing with trapped ions},\ }\href {https://doi.org/10.1098/rsta.2003.1205} {\bibfield  {journal} {\bibinfo  {journal} {Philosophical Transactions of the Royal
  Society of London. Series A: Mathematical, Physical and Engineering Sciences}\ }\textbf {\bibinfo {volume} {361}},\ \bibinfo {pages} {1349–1361} (\bibinfo {year} {2003})}\BibitemShut {NoStop}%
\bibitem [{\citenamefont {Manakov}\ \emph {et~al.}(1986)\citenamefont {Manakov}, \citenamefont {Ovsiannikov},\ and\ \citenamefont {Rapoport}}]{manakov_1986}%
  \BibitemOpen
  \bibfield  {author} {\bibinfo {author} {\bibfnamefont {N.~L.}\ \bibnamefont {Manakov}}, \bibinfo {author} {\bibfnamefont {V.~D.}\ \bibnamefont {Ovsiannikov}},\ and\ \bibinfo {author} {\bibfnamefont {L.~P.}\ \bibnamefont {Rapoport}},\ }\bibfield  {title} {\bibinfo {title} {Atoms in a laser field},\ }\href {https://doi.org/https://doi.org/10.1016/S0370-1573(86)80001-1} {\bibfield  {journal} {\bibinfo  {journal} {Physics Reports}\ }\textbf {\bibinfo {volume} {141}},\ \bibinfo {pages} {320} (\bibinfo {year} {1986})}\BibitemShut {NoStop}%
\end{thebibliography}%

\end{document}